# The origins of electromechanical indentation size effect in ferroelectrics

M. Gharbi[1], Z.H. Sun[1], P. Sharma[1,2,*], K. White[1]

[1]Department of Mechanical Engineering,
[2]Department of Physics
University of Houston, Houston, TX, 77204, U.S.A **Abstract**----Metals exhibit a size-dependent hardening when subject to indentation. Mechanisms for this phenomenon have been intensely researched in recent times.[1-4] Does such a size-effect also exist in the electromechanical behavior of ferroelectrics?---if yes, what are the operative mechanisms?  Our experiments on BaTiO$_3$ indeed suggest an electromechanical size-effect. We argue, through theoretical calculations and differential experiments on another non-ferroelectric piezoelectric (Quartz), that the phenomenon of *flexoelectricity* (as opposed to dislocation activity) is responsible for our observations. Flexoelectricity is the coupling of strain gradients to polarization and exists in both ordinary and piezoelectric dielectrics. In particular, ferroelectrics exhibit an unusually large flexoelectric response.

The indentation size effect of hardness in metals is generally attributed to dislocation activities. The model developed by Nix and Gao[3] argues the role of geometrically necessary dislocations associated with the strong strain gradient characteristically located below indenters. Nix and his colleagues have also conducted uniaxial compression experiments on micro/nano-pillars of metals, demonstrating the pillar size dependence of strength.[5-7] However, since no strain gradient is involved in these experiments, a different mechanism, dislocation starvation hardening, was proposed,[6] although the exact origins are still an active area of research.[8] In ferroelectrics also, increasing hardness with decreasing indenter radius has been observed for both PZT as well as BaTiO$_3$.[9,10] Although, Schneider and his colleagues reported the elastic modulus to be independent of indenter radius, the absence of contact stiffness versus contact radius curves for each indenter limits any direct comparison with our work. In the present work we pay careful attention to how the elastic behavior changes in ferroelectrics as a function of indent size.  Although the elastic properties of ordinary metals and ceramics[11] are nearly size-independent down to a few nanometers, ferroelectrics[12] and certain amorphous materials[11] may prove exception to the rule.

The so-called *elastic* indentation size-effect is nicely illustrated by our experiments on BaTiO$_3$ (001) oriented single crystals ($5 \times 5 \times 1$ mm).  A series of nanoindentations with a Berkovich indenter provided the contact stiffness versus contact radius curve (s-a curve) for comparison with the theoretically computed s-a curve for a circular flat indenter of various indenter radii. Considering the geometry-independent stiffness and contact radius relationship in the case of purely mechanical loading, the s-a curve for Berkovich indenter can be experimentally obtained in a single experiment. Since it is difficult to manufacture and maintain a conical indenter with a sharp tip, the reliable data in the small-

---

scale can not be obtained. Therefore, we adopted a sharp Berkovich indenter (three-sided pyramid, the tip radius ~50 nm) at the expense of well-defined contact radius. The area function of the indenter tip, $A = f(h_c)$ was carefully calibrated using standard procedures,[13] where $h_c$ is the contact depth. Although the projected contact area is not circular for the Berkovich indenter, the effective contact radius is calculated from the contact area by $\pi a^2 = f(h_c)$ and this approximation is quite good for small depths.

On the theoretical side, Karapetian el al.'s[14] have provided a detailed model of piezoelectric indentation. Within the assumption of transverse isotropy and restriction of indenter shape to be a cylinder, they interrelate applied concentrated force $P$, concentrated charge $Q$, indentation depth $w$ and tip potential $\psi_0$:

$$P = \frac{2aC_1^* w}{\pi} + \frac{2aC_3^* \psi_0}{\pi}$$
$$Q = \frac{2aC_2^* w}{\pi} + \frac{2aC_4^* \psi_0}{\pi}$$
(1)

The contact stiffness, under purely mechanical loading is: $s = 2\frac{C_1^*}{\pi} a$

We note here that $C_1^*$ is not just the elastic modulus but a combination of elasticity, piezoelectric and dielectric tensor components. The important aspect to keep in mind is that contact stiffness varies linearly with "a" or alternatively the ratio of contact stiffness to contact radius is a size-independent "constant". We note that a similar relation is obtained for indentation on elastically isotropic half spaces,[13,15] $C_1^*$ is $\pi E_r$, where $E_r$ is the reduced elastic modulus. This stiffness relation was derived for an axisymmetric indenter (circular contacts), however, it has been shown that it works well even for the non-axisymmetric shapes, provided a small correction factor is used.[13,15]

In Figure (1) we plot both our experimental results as well as the results from the aforementioned model based on classical piezoelectricity.

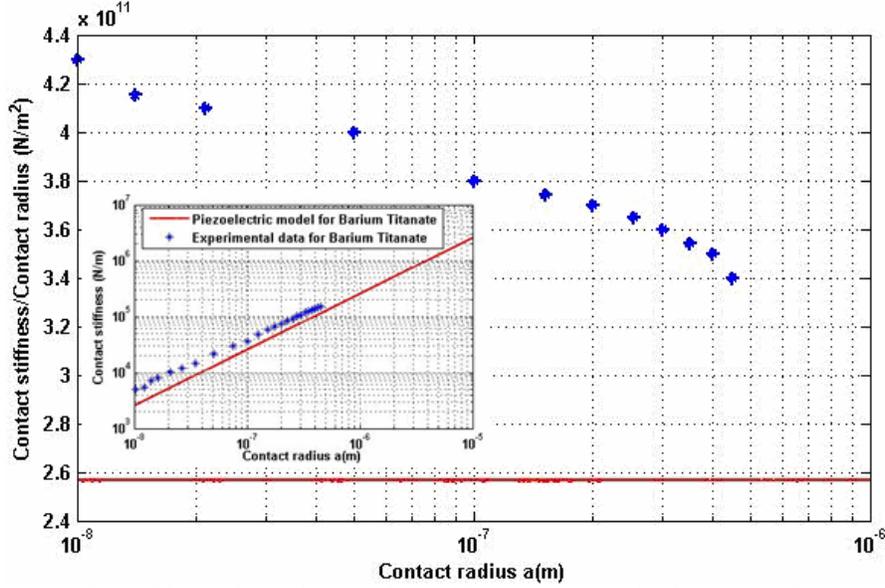

Figure 1: Variation of the ratio of the contact stiffness and contact radius with respect to the contact radius $a$ for BaTiO$_3$. The inset graph plots the contact stiffness versus radius.

From Figure (1) two points are well-evident: (i) our experiments suggest a strong indentation size-effect. (ii) classical piezoelectricity, as anticipated from the model based on classical piezoelectricity[14], fails to capture this. We may be tempted to resort to dislocations based arguments to explain this size-effect. Our theoretical analysis (to follow) however hints towards another explanation, namely, *flexoelectricity*.

Piezoelectricity requires that the crystalline unit cell lack centro-symmetry (for example NaCl is not piezoelectric while BaTiO$_3$ is). An underappreciated fact is that in the presence of inhomogeneous strain, inversion symmetry can be broken leading to the development of polarization even in non-piezoelectric materials.

$$(P)_i = (d)_{ijk}(\varepsilon)_{jk} + (f)_{ijkl}\nabla_l(\varepsilon)_{jk} \qquad (2)$$

where $d$ is the third order piezoelectric tensor and $f$ is the fourth order flexoelectric tensor. This is well illustrated for graphene through *ab initio* simulation under bending (which is manifestly non-piezoelectric)---see Figure (2 for graphene).

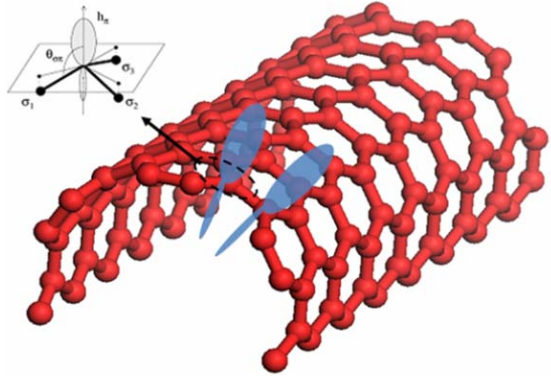

Figure 2: Bending of a graphene sheet. A non uniform strain (bending) of a readily non piezoelectric two dimensional graphene sheet causes a rehybridization of π and σ orbitals. For simplicity, only the asymmetric orthogonal π-hybrids (π and σ mixing denoted by $h_\pi$) are represented in the figure. Each $\sigma_i$ bond (between nearest neighbors) is deflected down by an angle (with respect to its plane configuration (see light black lines in the insert)). The $\sigma_i$ bonds are now in pyramidal shape (see dark black lines in the insert) and form an angle $\theta_{\sigma\pi}$ with the $h_\pi$ hybrids. As a result, the center of electronic charge at each atomic site is displaced outwards from the nuclear charge inducing a net polarization.[16]

Recently one of us has clarified some of the basic mechanisms behind flexoelectricity as well as evaluated flexoelectric properties through atomistic calculations[17]. Other very interesting works have also appeared. Experiments on finding flexoelectric properties were pioneered by Cross and co-workers[18] who have established that flexoelectric constants are three orders of magnitude larger than ordinary dielectrics. Zubko[19] have recently published the experimental characterization of the complete flexoelectric tensor of SrTiO$_3$. Recently, we have shown the prospects of enhanced piezoelectricity in nanostructures[12] due to flexoelectricity, its role in the origins of the dead-layer in ferroelectric based nanocapacitors[20] and underscored Cross's idea[21] of the possibility of creating apparently piezoelectric materials without using piezoelectric materials[22].

In prior work we have presented a mathematical formulation of the theory of flexoelectricity[23]. The equations are quite complicated even for an isotropic continuum let alone a anisotropic crystal which also exhibits direct piezoelectricity (as is the case for BaTiO$_3$). Nevertheless we, employing a perturbation approach coupled with guidance from some partial numerical calculations, have been successful in generating closed form expression for the effect of flexoelectricity on indentation. Details of the model itself will be presented elsewhere. To summarize, we find the following (for purely mechanical loading):

$$P = \frac{2a}{\pi} C_1^* w - \frac{2}{\pi a} f_1^* w \frac{\varepsilon}{(A^i)^2} \left( \varepsilon e^{-\frac{A^i}{\varepsilon} a} - \varepsilon + A^i a \right) \qquad (3)$$

The ratio of the contact stiffness to the contact radius is then:

$$\frac{s}{a} = \frac{1}{a} \frac{\partial P}{\partial w} = \frac{2}{\pi} C_1^* - \underbrace{\frac{2 f_1^*}{\pi a^2} \frac{\varepsilon}{(A^i)^2} \left( \varepsilon e^{-\frac{A^i}{\varepsilon} a} - \varepsilon + A^i a \right)}_{size-effect} \qquad (4)$$

where $A^i$ and $f_1^*$ are constants depending on material properties and $\varepsilon$ is the approximate value of flexoelectric and piezoelectric ratio. Equation (3) shows the presence of a size-effect due to flexoelectricity (underlined terms).

Experiments of Cross[18] indicate that flexoelectric coefficients are quite large for $BaTiO_3$ and thus we anticipate flexoelectricity to contribute significantly. However, can other mechanisms such as dislocation hardening be ruled out? While the latter is a complex problem, two simple steps (one qualitative argument and the other a simple differential experiment) can help provide some insight. Quartz is not expected to exhibit a *widely* different dislocation activity from $BaTiO_3$, on the other hand, its flexoelectric properties are nearly 3 orders of magnitude lower than that of $BaTiO_3$. Accordingly, we also carried out an identical set of indentation experiments on Quartz following essentially the same procedure as that for $BaTiO_3$. Comparisons of our flexoelectricity based model with both our experiments are shown in Figure (3).

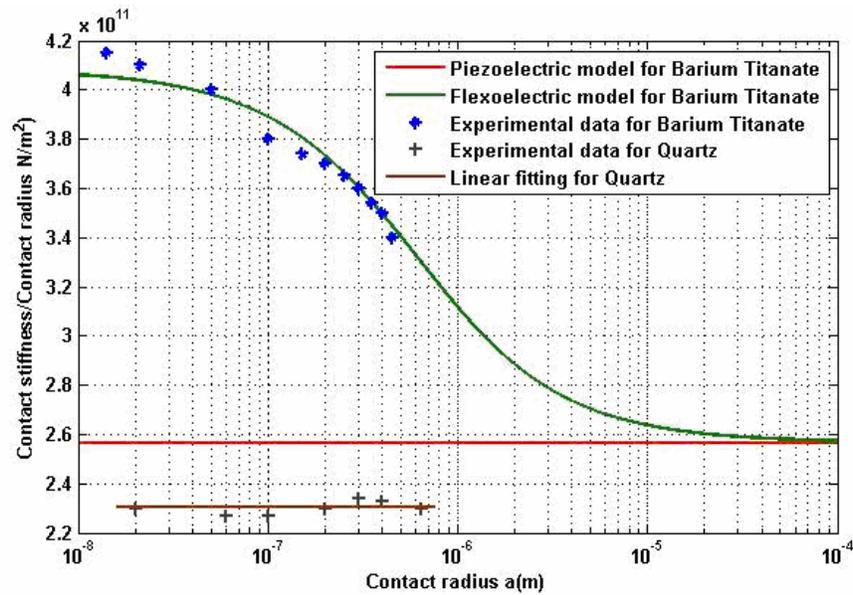

Figure 3: Variation of the ratio of the contact stiffness and contact radius as a function of contact radius. Experimental results for both $BaTiO_3$ and Quartz are plotted along with the results from the flexoelectricity and classical piezoelectricity based models.

As well-evident from Figure (3), the agreement between our model and experiments is quite close---remarkable since there is no "fitting or calibration" performed between our theoretical/computational model and experiments. Quartz shows no size-effect in the experiments, and is clearly anticipated by our model since the flexoelectric constants are quite small for this material (i.e. the underlined term in Equation 3 is essentially zero in the range of contact sizes shown in Figure 3).

The role of mechanisms such as domain wall activity, dislocation based mechanisms among others to explain the observed electromechanical size-effect cannot be conclusively ruled out. However, our comparative experiments on both $BaTiO_3$ and Quartz as well as the close agreement of our model with experiments indicates that flexoelectricity is most likely the dominant mechanism behind the observed size-effect.